\RequirePackage{fix-cm}
\documentclass[twocolumn,epjc3]{svjour3}  
\smartqed  
\RequirePackage{graphicx}
\RequirePackage[numbers,sort&compress]{natbib}
\usepackage{booktabs,verbatim} 
\usepackage[usenames,dvipsnames]{color}
\usepackage{multirow,bigdelim}
\usepackage{ulem}
\usepackage{color}
\usepackage{textcomp} 
\usepackage{tabularx}
\usepackage{amsmath}
\begin{document}

\title{Long-term evolution of the neutron rate at the Canfranc Underground Laboratory}

\author{S.~E.~A.~Orrigo\thanksref{e1,IFIC} \and	J.~L.~Tain\thanksref{IFIC} \and	N.~Mont-Geli\thanksref{UPC} \and A.~Tarife{\~n}o-Saldivia\thanksref{UPC} \and L.~M.~Fraile\thanksref{Madrid} \and	M.~Grieger\thanksref{Dresden} \and J.~Agramunt\thanksref{IFIC} \and	A.~Algora\thanksref{IFIC} \and D.~Bemmerer\thanksref{Dresden} \and F.~Calvi{\~n}o\thanksref{UPC} \and	G.~Cort{\'e}s\thanksref{UPC} \and A.~De~Blas\thanksref{UPC} \and I.~Dillmann\thanksref{Triumf} \and A.~Dom{\'i}nguez~Bugar{\'i}n\thanksref{Madrid} \and R.~Garc{\'i}a\thanksref{UPC} \and E.~Nacher\thanksref{IFIC} \and A.~Tolosa-Delgado\thanksref{IFIC}}

\thankstext{e1}{e-mail: Sonja.Orrigo@ific.uv.es}

\authorrunning{S.~E.~A.~Orrigo $et~al.$} 

\institute{Instituto de F{\'i}sica Corpuscular, CSIC-Universidad de Valencia, E-46071 Valencia, Spain \label{IFIC}
   \and
    Institute of Energy Technologies (INTE), Technical University of Catalonia (UPC), E-08028 Barcelona, Spain \label{UPC}
   \and
	  Grupo de F{\'i}sica Nuclear \& IPARCOS, Universidad Complutense de Madrid, E-28040 Madrid, Spain \label{Madrid}
   \and
	  Helmholtz-Zentrum Dresden-Rossendorf (HZDR), 01328 Dresden, Germany \label{Dresden}
	 \and
	  TRIUMF, 4004 Wesbrook Mall, Vancouver, British Columbia V6T 2A3, Canada \label{Triumf}
}

\date{Received: date / Accepted: date} 

\maketitle

\begin{abstract}
We report results on the long-term variation of the neutron counting rate at the Canfranc Underground Laboratory, of importance for several low-background experiments installed there, including rare-event searches. The measurement campaign was performed employing the High Efficiency Neutron Spectrometry Array (HENSA) mounted in Hall~A and lasted 412 live days. The present study is the first long-term measurement of the neutron rate with sensitivity over a wide range of neutron energies (from thermal up to 0.1~GeV and beyond) performed in any underground laboratory so far. Data on the environmental variables inside the experimental hall (radon concentration, air temperature, air pressure and humidity) were also acquired during all the measurement campaign. We have investigated for the first time the evolution of the neutron rate for different energies of the neutrons and its correlation with the ambient variables.
\end{abstract}

\section{Introduction}
\label{intro}

Underground (UG) laboratories provide a very-low background environment because the muon flux coming from cosmic rays is largely suppressed by the rock overburden. Hence they are well suited to host a vast spectrum of experiments characterized by extremely low event rate, ranging from nuclear astrophysics to astroparticle physics experiments.

The Canfranc Underground Laboratory (LSC) \cite{Ianni_2016} is located 850 m under the Mount Tobazo in the Aragonese Pyrenees in Spain. A recent measurement of the muon flux at LSC indicates a reduction of a factor 10$^5$ in comparison to surface (2500 m.w.e.) \cite{LSC2}. LSC hosts a number of experiments investigating fundamental questions of modern physics \cite{LSC_2020}, such as the nature of dark matter (ANAIS~\cite{ANAIS_2019}, ArDM~\cite{ArDM_2017}, TREX-DM~\cite{TREX_2020}), neutrino properties beyond the Standard Model (NEXT~\cite{NEXT_2018}, CROSS~\cite{CROSS_2020}) and the origin of the chemical elements in the Universe (the proposed CUNA project~\cite{CUNA}).

Neutrons constitute a main limitation for experiments dealing with rare-event searches. Even if the part of neutron flux produced by cosmic-ray muons is largely suppressed in UG laboratories, natural radioactivity is not. Radiogenic neutrons still originate in the rocks and cavity walls by ($\alpha$,n) reactions, induced by the $\alpha$ radioactivity of the $^{238}$U and $^{232}$Th decay chains on light elements, and the spontaneous fission of $^{238}$U. Neutrons have a large penetrability because they are not affected by the Coulomb barrier, so they can induce background signals in the detectors. For example, the elastic scattering of neutrons can mimic the Weakly Interacting Massive Particles (WIMPs) signals, while the inelastic scattering or neutron-capture processes produce $\gamma$ rays which can influence neutrinoless $\beta\beta$ decay searches. Moreover, since the neutron cross section critically depends on neutron energy, a different energy distribution can have a drastic effect on the various experiments. Such as, low-energy neutrons constitute a critical background for experiments investigating nucleosynthesis processes \cite{Tain_2016}, while high-energy neutrons affect dark matter experiments since polyethylene and water shields are not capable to moderate them completely \cite{Aprile_2013}. A spectroscopic measurement of the neutron rate is therefore crucial. Furthermore, information on the long-term behaviour of the neutron background is of relevance for the UG experiments, but poorly known at present. A seasonal variation of the neutron rate can be important especially for experiments looking for the annual modulation of dark matter \cite{ANAIS_2019}.

When ignoring different depths, variations of the neutron background at different UG laboratories are due to differences in the geological composition of the rock as well as in the amount of uranium and thorium in the environment and in the water content of the rocks surrounding the experimental setup \cite{Best_2016}. Local rock composition can vary even for different locations of the same UG laboratory (e.g., between Halls~A and C at the Laboratori Nazionali del Gran Sasso (LNGS) \cite{Wulandari_2004}). Moreover, the concentration of the inert and radioactive gas radon close to the setup, strongly dependent on the air circulation and the eventual presence of a forced ventilation system, can also influence the neutron background locally \cite{Carson_2004}. $^{222}$Rn (with half-life $T_{1/2}$ = 3.8(3) days) originates from the $^{238}$U decay chain, can travel with air for long distances and penetrates in porous materials (soil, sedimentary rock, concrete, etc.). When $^{222}$Rn or its descendants decay inside those materials, they may produce fast neutrons by ($\alpha$,n) reactions that can escape into the experimental room \cite{Stenkin_2017}.

Therefore it is of paramount importance to measure and fully characterize the neutron flux at the precise location where experiments are performed, and to do it as a function of neutron energy. Since the neutron background in UG laboratories is low, measurements have a low rate and last for months, demanding detectors with high efficiency for neutron detection, high background discrimination capability and long-term stability. These requirements are fulfilled by moderated $^3$He proportional counters \cite{Thomas_2012,Jordan,Tain_2016,Grieger_2020}. In particular, Ref.~\cite{Jordan} reports a first pioneering measurement carried out in the (almost empty) Hall~A of LSC employing six moderated $^3$He detectors and lasting 26 days.

In October 2019 we started a long-term measurement of the neutron flux in the Hall~A of LSC, using the High Efficiency Neutron Spectrometry Array (HENSA) \cite{HENSA1}. HENSA is composed of long proportional counters filled with $^3$He gas embedded in high-density polyethylene (PE) moderators with different thickness, achieving sensitivity to neutrons with energies ranging from thermal to 0.1~GeV and beyond. The aim of the measurement is trifold: to measure the integral neutron flux in Hall~A at LSC, to study the long-term evolution of the neutron rate looking for possible seasonal variations and to characterize the energy spectrum of the neutron background precisely.

In this paper we present new results on the long-term evolution of the neutron counting rate, observed in a period of 412 live days and analysed in connection to the measurements of the environmental variables at LSC (radon concentration, air temperature, air pressure and humidity). 

A limited number of studies show the time evolution of the thermal neutron rate in UG laboratories, employing $^6$LiF+ZnS(Ag) scintillators with sensitivity to the thermal range \cite{Alekseenko_2010,Alekseenko_2017,Stenkin_2017}. The present study is, to our knowledge, the first long-term measurement carried out with a spectrometer composed of moderated $^3$He counters with sensitivity over a wide range of neutron energies. This feature, together with the excellent stability of the HENSA spectrometer, allows us to investigate the long-term behaviour of the neutron rate for different neutron energies for the first time. As discussed, this measurement is of importance for a number of experiments characterized by low event rate at LSC.

The paper is organized as follows. Section~\ref{exp} describes the experimental setup and the measurement campaign performed in Hall~A. The details of data analysis are given in Section~\ref{an}. The experimental results are presented and discussed in Section~\ref{results}. In particular, the total neutron count rates are given in Section~\ref{nRate}, where they are also compared to the results from Ref.~\cite{Jordan}. The long-term evolution of the neutron rate is shown for the different energy ranges in Section~\ref{evol}. Section~\ref{envr} presents the time evolution of the ambient data and discuss their possible correlation with the neutron rate. Finally, Section~\ref{concl} summarises our conclusions.

\section{\label{exp}The experiment}

A long-term measurement of the neutron flux has been carried out at the Hall~A of LSC employing the HENSA spectrometer to detect the ambient neutrons. The Hall~A measurement campaign started in October 2019 and ended in March 2021.

\subsection{\label{setup}Experimental setup}

\begin{table*}[!th]
	\caption{Characteristics of the PE moderators and covering materials surrounding the one-inch diameter cylindrical $^3$He proportional counters ($P$ = 10 atm, $l$ = 60 cm) of the HENSA spectrometer in the measurement at the Hall~A of LSC.	The total size of each PE block is shown in the second column, while the third column reports the thickness of the additional material. The typical neutron energy sensitivity range [$E_\text{i}$,$E_\text{f}$] for each detector, estimated as the energy interval where the simulated response is between 10\% and 90\% of the corresponding accumulated response, is given in the last two columns.}
	\label{Tab1}
	\centering
	  \begin{tabular}{lllll}
		  \hline
 		  Detector & PE size (cm$^3$) & Additional material & \multicolumn{2}{c}{Neutron energy range (MeV)} \\ \hline
			   &              &                      & $E_\text{i}$ & $E_\text{f}$ \\
			E1 & No moderator &                      & 3$\times$10$^{-10}$           & 10$^{-6}$ \\
			E2 & 4.5 $\times$ 4.5 $\times$ 70 &      & 4$\times$10$^{-9}$           & 4$\times$10$^{-3}$ \\
			E3 & 7 $\times$ 7 $\times$ 70 &          & 6$\times$10$^{-8}$   & 4$\times$10$^{-1}$ \\						
			E4 & 12 $\times$ 12 $\times$ 70 &        & 10$^{-6}$ & 6 \\
			E5 & 18 $\times$ 18 $\times$ 70 &        & 6$\times$10$^{-5}$            & 6$\times$10$^{1}$ \\
			E6 & 22.5 $\times$ 22.5 $\times$ 70 &    & 3$\times$10$^{-3}$                  & 8$\times$10$^{2}$ \\			
			E7 & 27 $\times$ 27 $\times$ 70 &        & 10$^{-1}$                    & 2$\times$10$^{3}$ \\					
			E8 & 21 $\times$ 21 $\times$ 70 & 5 mm Pb & 4$\times$10$^{-3}$                 & 4$\times$10$^{3}$ \\						
		  E9 & 25 $\times$ 25 $\times$ 70 & 10 mm Pb, 0.7 mm Cd & 3       & 6$\times$10$^{3}$ \\		
		  \hline								
		\end{tabular}
\end{table*}

\begin{figure}[!t]
  \centering
	\includegraphics[width=1\columnwidth]{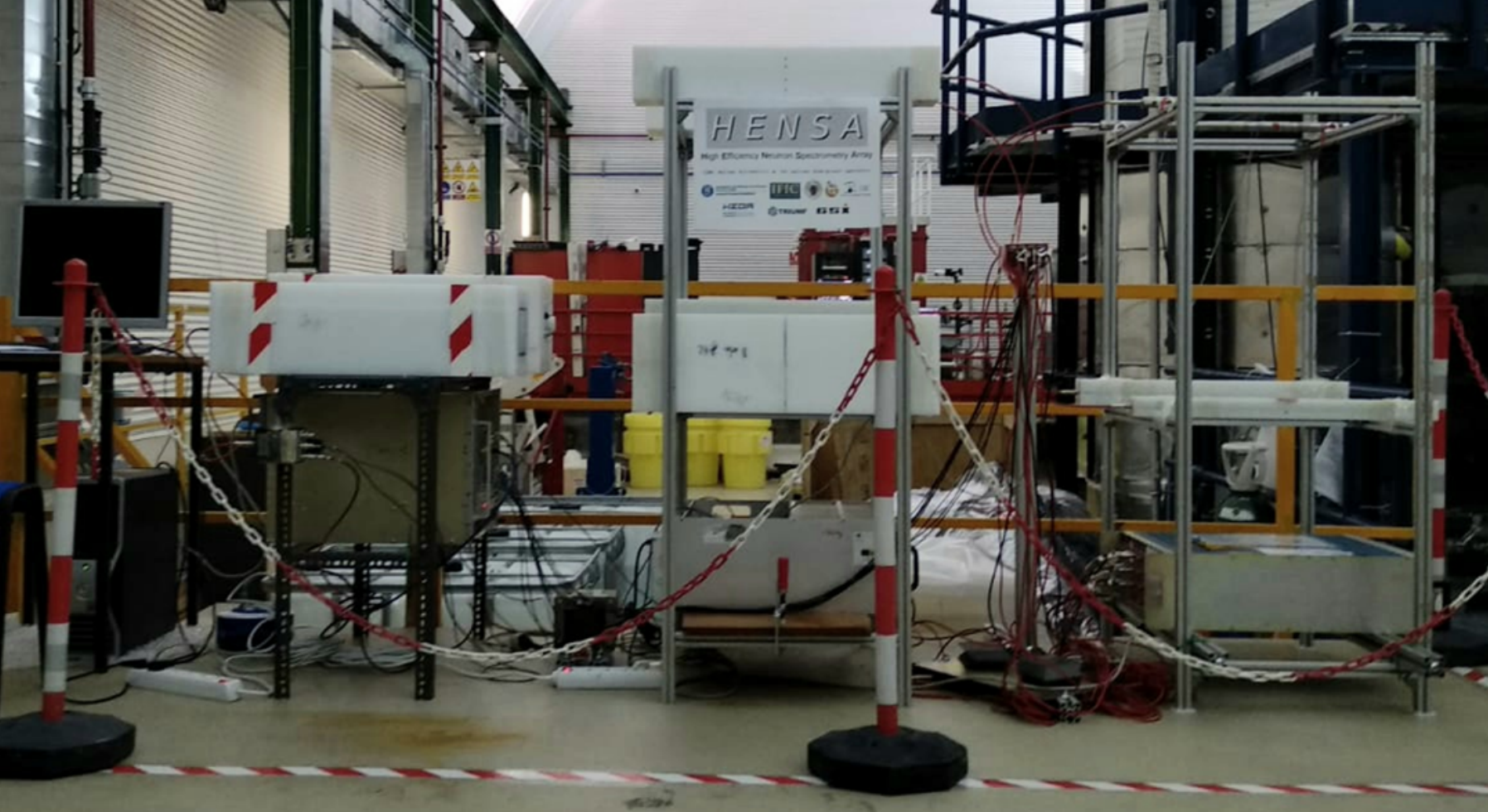}
 	\caption{The High Efficiency Neutron Spectrometry Array (HENSA) mounted in the Hall~A at the Canfranc Underground Laboratory.}
  \label{Fig1}
\end{figure}

HENSA \cite{HENSA1}, which is based on the Bonner Spheres Spectrometer (BSS) principle \cite{Thomas_2012}, is an array of nine one-inch diameter cylindrical proportional counters filled with $^3$He gas at a pressure $P$~=~10~atm and having an active length $l$~=~60~cm (Fig.~\ref{Fig1}). The tube walls are made of stainless steel providing in general much lower internal background than aluminum \cite{Crane1991}. The detection of thermal or moderated neutrons is achieved in the $^3$He tubes using the $^3$He(n,p)$^3$H reaction to convert the neutrons into charged particles (p and $^3$H) which are then detected, producing a spectrum of deposited energy with a shape that is characteristic of $^3$He counters (see Section~\ref{an}).

HENSA has been carefully designed to provide very good sensitivity over a large energy range. Since the neutron detection efficiency depends on both the neutron energy and moderator thickness \cite{Thomas_2012,Jordan}, in order to achieve sensitivity for different neutron energies (from thermal to the GeV range) and to increase the sensitivity at both extremes of the energy range, each tube of HENSA is embedded in a matrix of different and carefully-selected materials. These materials, listed in Table~\ref{Tab1}, include: PE of different thickness, acting as a neutron moderator; cadmium, which is a strong absorber of thermal energy neutrons; lead, acting as a converter for high-energy neutrons through (n,$x$n) reactions (where $x$ indicates the multiplying factor for neutrons), providing sensitivity above 20 MeV. A bare $^3$He tube is also included to enhance the sensitivity to thermal neutrons, which do not require moderation. The detectors are ordered in Table~\ref{Tab1} according to the sensitivity to increasing neutron energy. The use of long $^3$He tubes provides HENSA with neutron efficiency around one order of magnitude \cite{HENSA1} larger than standard BSS \cite{Thomas_2012} over the full range of neutron energies (from 5 to 15 times higher depending on the particular value of neutron energy considered).

The HENSA response to neutrons was determined by Monte Carlo simulations performed using both the FLUKA \cite{Grieger_2020} and GEANT4 \cite{HENSA1} codes, with excellent agreement. The reader can find an example of such a response in Fig.~2 of Ref.~\cite{Orr_2021}, reporting the results of a first simulation including each detector separately and performed up to 10$^{4}$ MeV. The last two columns of Table ~\ref{Tab1} report the typical neutron energy sensitivity range for each detector, obtained from these simulations considering the energy interval where the response is between 10\% and 90\% of the corresponding accumulated response. Improvements of the simulations are ongoing such as the inclusion of the detector cabinet, floor, etc. Further details on the spectrometer development and the Monte Carlo simulations of its response will be provided in a forthcoming publication \cite{HENSA}. Knowing the HENSA response, which depends on the efficiency of each detector as a function of neutron energy, a deconvolution procedure allows to combine the neutron rates measured by every counter to deduce both the absolute value and spectral shape of the neutron flux \cite{Thomas_2012,Jordan,Grieger_2020,Orr_2021}.

HENSA employs nuclear digital electronics. Each detector is connected to a charge-sensitive preamplifier (CAEN A1422 and Canberra 2006 modules). A CAEN SY127 High-Voltage (HV) power supply, equipped with the A333 and A331 HV modules, feeds each preamplifier channel. The preamplifier output signal is sent to the data acquisition system that uses a SIS3316 VME module, a 16-channels sampling digitizer from Struck Innovative Systems \cite{STRUCK} with 250 MHz sampling frequency and 14 bit resolution. The acquisition is controlled by the self-triggered GASIFIC data acquisition system (DACQ) developed at IFIC-Valencia \cite{Agr_2016}. The software provides an online acquisition control and visualization, allowing the online analysis while the raw data is written to disk. A trapezoidal Finite Impulse Response (FIR) filter produces a short waveform for discrimination purposes (fast filter). Each signal out of the fast filter above a given threshold is associated with a time reference (\textit{time~stamp}) of 4 ns resolution and generates an internal trigger for processing the input pulse. The latter is processed with a second trapezoidal FIR filter (slow filter) for precise determination of the waveform amplitude, which is proportional to energy. Furthermore, the entire waveform is recorded for a time interval of 20 \textmu s, which allows pulse shape discrimination (PSD) of spurious signals if needed \cite{Beltran_2011,Langford_2013}. The full raw data from each detector is then processed with a specific sorting program which generates TTrees of the ROOT data analysis framework \cite{ROOT} to be used for the offline data analysis.

\subsection{\label{measures}Measurement campaign in Hall~A}

The HENSA spectrometer was mounted in Hall~A as shown in Fig.~\ref{Fig1}. The $^3$He proportional counters, embedded in PE moderators and absorber or multiplier materials (see Table~\ref{Tab1}), were placed on a set of light structures. This placement took into account electronic noise considerations, demanding short cable lengths.

The measurements in Hall~A with the full setup (\textit{Phases}~1 and 2) lasted from October~3, 2019 to July~23, 2020 for a total of 248 live days, demonstrating the excellent stability of the setup. During \textit{Phase}~2 (135 live days) a 20-atm tube replaced the 10-atm tube in E9, in order to assess the systematic effects due to a different pressure \cite{HENSA}. Afterwards, from August~1, 2020 to March~4, 2021 (\textit{Phase}~3, 164 live days) the thermal and epithermal detectors of the setup (E1 and E2 in Table~\ref{Tab1}, respectively) were the only ones left measuring in Hall~A. One part of the other detectors was employed in measurements of the neutron flux above ground and the other part was moved to the Hall~B of LSC to start a measurement of the neutron background close to ANAIS \cite{Mon_2021}.

During the Hall~A campaign, the data acquisition was running continuously and controlled remotely on a daily basis. Besides short periods with intervention on the HENSA setup, time periods when disturbing activity was present in Hall~A (such as calibrations with sources performed by the other experiments) were excluded in the offline analysis and thus are not included in the above-mentioned live days. The amplitude response to thermal neutrons of each tube composing the HENSA spectrometer was characterized with $^{252}$Cf spontaneous fission neutron sources at the beginning of \textit{Phase}~1 and then at the end of both \textit{Phases}~2 and 3, showing no significant changes during the whole measurement campaign. This confirms the excellent stability over time of the $^3$He detectors.

The environmental conditions at LSC are constantly monitored in various positions inside the experimental halls. We have been provided with the ambient data from the AlphaGuard monitor located in Hall~A close to HENSA. The monitored variables are the $^{222}$Rn concentration, air temperature $T$, air pressure $P$ and relative humidity $H$. The instrument provides data averaged over a period of 10 minutes. 

\section{\label{an}Data analysis}

The present section describes the procedures employed for the analysis of the neutron data acquired in Hall~A. HENSA performed remarkably well during the entire measurement campaign. 

As anticipated in Section~\ref{setup}, neutrons are detected indirectly in the HENSA setup, through the detection of the proton and $^3$H particles produced (with kinetic energies of 573 and 191 keV, respectively) in the reaction of the incoming thermal or moderated neutron with the $^3$He gas filling of a given proportional counter. Complete absorption of both the proton and triton energies gives a signal peaked at the $Q$-value of the reaction (764 keV). However, due to the finite volume of the counters, it is possible that one or both reaction products are not fully stopped inside the sensitive volume, thus they only deposit a fraction of their energy in the detector (\textit{wall~effect}). This mechanism gives a characteristical shape to the spectrum with two steps at 191 and 573 keV, corresponding to complete detection of the triton energy with partial deposition of the proton energy inside the tube and vice versa. 

Common backgrounds observed in the $^3$He counters \cite{Jordan,Tain_2016,Best_2016} are, on the low-energy side, $\gamma$ rays and electronic noise. On the high-energy side, $\alpha$ radioactivity from the decay of uranium and thorium present in the counter walls generates background signals extending up to 9 MeV \cite{Has_1998} and at energies below the neutron peak, as well. High-voltage micro-discharges \cite{Heeger_2000} can contribute to the background in the whole range of interest. Our procedure to disentangle the different contributions is explained below.

\begin{figure}[ht!]
	\begin{minipage}{1.0\linewidth}
	  \centering
	  \includegraphics[width=1\columnwidth]{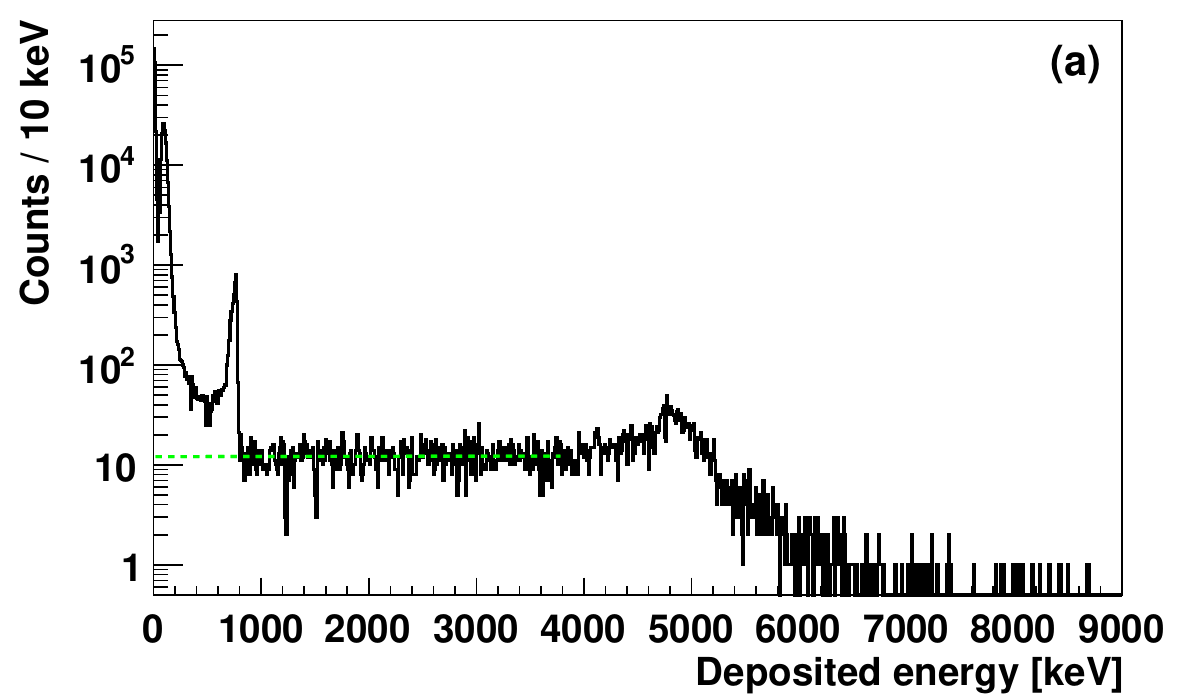}
	\end{minipage}
	\begin{minipage}{1.0\linewidth}
	  \includegraphics[width=1\columnwidth]{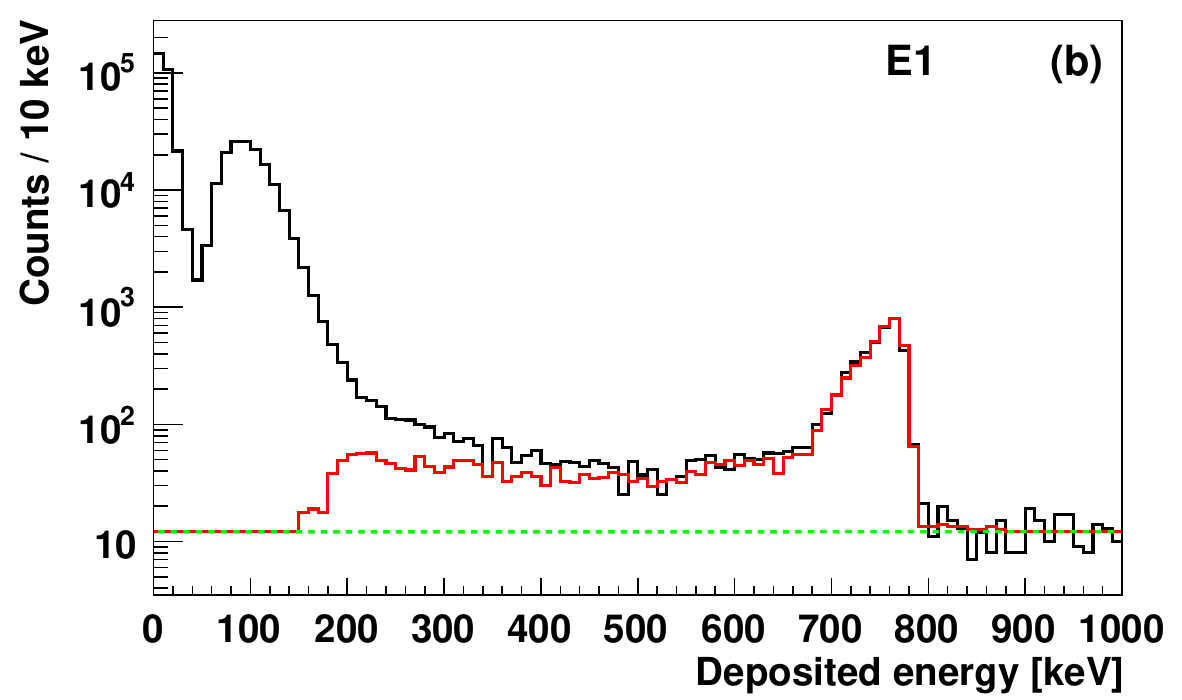}
	\end{minipage}
 	\caption{Example of a typical spectrum measured in a $^3$He proportional counter. The black histogram is the spectrum measured by the thermal detector E1 in Hall~A during \textit{Phase}~1. The green dashed line indicates the linear $\alpha$ background. (a) The region [$0,9000$] keV is shown. (b) The region [$0,1000$] keV is shown. The red histogram is the result of a measurement with a $^{252}$Cf source, on top of the linear $\alpha$ background (green dashed line) and fitted to the black histogram. Details are given in the text.}
  \label{Fig2}
\end{figure}

\begin{figure*}[th!]
  \centering	
	\includegraphics[width=2\columnwidth]{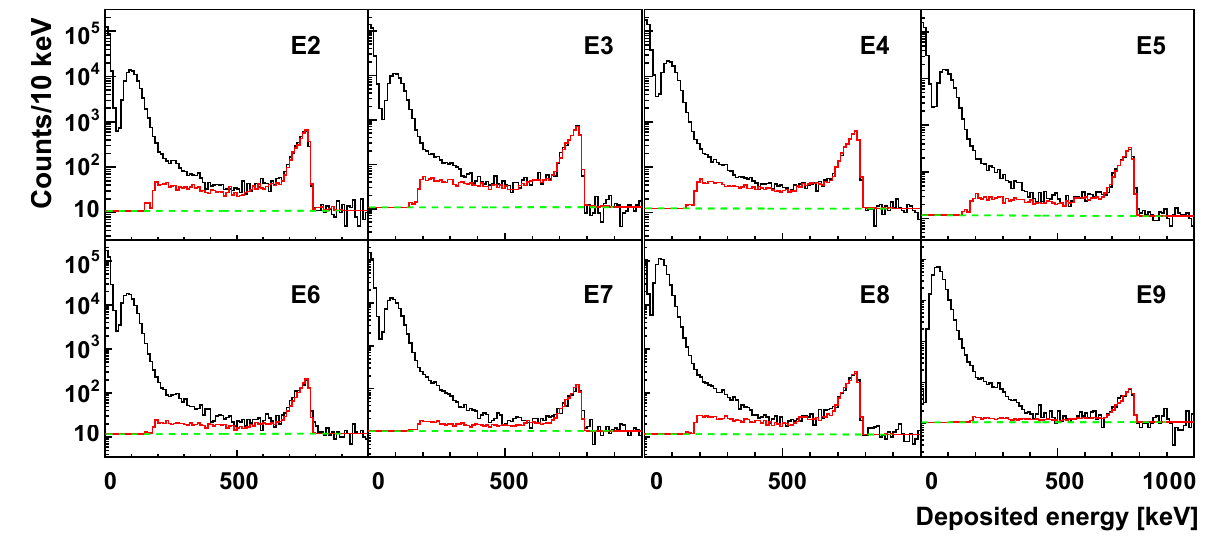}
 	\caption{Spectra measured in the $^3$He proportional counters from E2 to E9 (as listed in Table~\ref{Tab1}). The black histograms are the spectra measured by each detector in Hall~A during \textit{Phase}~1. The green dashed lines indicate the fit of the linear $\alpha$ background. The red histograms are the result of a measurement with a $^{252}$Cf source, on top of the $\alpha$ background and fitted to the black histograms, as done in Fig.~\ref{Fig1} for E1.}
  \label{Fig3}
\end{figure*}

The typical spectral shape from a $^3$He proportional counter is visible in Fig.~\ref{Fig2} which shows, as an example, the spectrum measured in the thermal detector E1 during the whole \textit{Phase}~1, covering 113 live days (black histogram). In Fig.~\ref{Fig2}a such a spectrum is shown up to 9 MeV. The neutron peak is visible at 764 keV. The high-energy component of the background, due to the $\alpha$ particles originating from the counter itself, peaks at around 4.8 MeV. A convenient way to account for such a background \cite{Jordan,Tain_2016,Best_2016} is a linear fit performed in the featureless region which is extrapolated to lower energies (green dashed line). In Fig.~\ref{Fig2}b the same spectrum is shown up to 1 MeV. The first peak visible at low energy is due to electrical noise, while the second and broader peak at around 100 keV arises from the $\gamma$ background. The neutron signals appear in the energy region [$150,900$] keV, as illustrated by the red histogram. The latter is the spectrum obtained in the E1 counter during the characterization with a moderated $^{252}$Cf neutron source, representing a \textit{pure} neutron response, summed to the linear $\alpha$ background (green dashed line) and fitted to the measured spectrum (black histogram). Similar spectra, obtained for the other HENSA detectors, are shown in Fig.~\ref{Fig3}.

Standard analysis procedures \cite{Jordan,Tain_2016,Best_2016,Grieger_2020} were used to independently analyse the spectrum measured by each counter and determine the signal count due to neutrons, $N_n$. For each detector, the $\alpha$ background was fitted in the region [$900,3800$] keV and extrapolated below (green dashed line in Fig.~\ref{Fig2}). The spectrum from the $^{252}$Cf neutron source is first converted into an unitary-area spectrum and then normalized (red histogram in Fig.~\ref{Fig2}b), taking into account the $\alpha$ background, to the measured spectrum (black histogram) by fitting the region of the neutron peak, i.e., [$730,800$] keV. Taking this energy region as a reference, the integral of the neutron peak in the measured spectrum, $N_\text{exp}$, can be expressed as:

\begin{equation}
  N_\text{exp} = N_\text{n} \cdot N_\text{0} + N_{\alpha} \,,
  \label{Eq1}
\end{equation}
\\
where $N_\text{0}$ and $N_{\alpha}$ are the integrals of the unitary-area $^{252}$Cf spectrum and $\alpha$ background in the same region, respectively. In this way one is able to determine $N_\text{n}$ which, together with the measurement live time, provides the neutron count rate. The systematic uncertainty of this procedure was estimated by varying the reference region of the neutron peak as well as the region used for the fit of the $\alpha$ activity and amounts to a few \% (1\% to 5\% depending on the detector). The procedure described above was employed to independently analyse the spectrum measured by each detector, separately for each one of the three measurement periods (\textit{Phases}~1, 2 and 3). Since in all the cases the signal to background ratio is high enough, the procedure provided a clean separation of neutron counts without the need to resort to PSD techniques, which have the added complication of determining accurately the efficiency of the applied cuts.

During \textit{Phase}~3 the setup was quite reduced, with only two detectors left (the thermal E1 and epithermal E2 ones), changing the amount of surrounding PE moderators which in turn may affect the neutron count rate. The amount of such an effect, estimated using the procedure described below, is within a 2\% systematic uncertainty which has been included for the \textit{Phase}~3 data.

The neutron rate observed in each detector $n_i$ arises from a combined effect of the detector response $\epsilon_{ij}$ and incident neutron flux $\phi_j$ for each energy bin $j$ \cite{Orr_2021}:

\begin{equation}
  n_i = \sum_j \epsilon_{ij} \phi_j \,.
  \label{Eq2}
\end{equation}

Therefore, we compared the neutron rate calculated in the detectors E1 and E2 for two different simulations of the detector response: the response of isolated detectors and that of detectors grouped in cabinets, in the same configuration as they were mounted in Hall~A during the measurement. Such a calculation of the neutron rate used various simulated neutron fluxes from Ref.~\cite{Mon_2021} representing different compositions of the rock and concrete in Hall~A, which are not known at present. The differences obtained in the rates of the detectors E1 and E2 with the two responses are within 2\%.

The resulting neutron count rates for each \textit{Phase} are presented in Section~\ref{nRate}, where they are also compared to the results of a previous measurement \cite{Jordan}. In order to investigate the possible correlation between the neutron count rate and the environmental data, the neutron rate data were grouped in time periods of one month and compared to the ambient data measured by the AlphaGuard monitor corresponding to the same periods. The long-term evolution of the neutron rate in the different energy ranges and the environmental variables are shown in Sections~\ref{evol} and \ref{envr}, respectively.

\section{\label{results}Experimental results}

\subsection{\label{nRate}Neutron counting rates}

The neutron count rates obtained for each detector and each measurement period (\textit{Phases}~1, 2 and 3) as explained in Section~\ref{an} are given in Table~\ref{Tab2} (columns 2 to 4), where the quoted uncertainties include also the systematics of the analysis procedure and the additional 2\% contribution in \textit{Phase}~3 (see Section~\ref{an}). The detectors are listed according to their sensitivity to increasing neutron energy, i.e., from thermal to high-energy neutrons. The detector E9 had a pressure of 10 and 20 atm in \textit{Phases}~1 and 2, respectively; thus to compare properly its rates of both periods we considered the responses simulated for E9 using the two different pressure values, obtaining a factor of 0.92(2) used to correct the E9 rate in \textit{Phase}~2.

When comparing the different \textit{Phases}, a variation is observed in the rates of the thermal (E1) and epithermal (E2) detectors while for the other detectors, sensitive to higher energy regions, the rates are essentially unchanged. The effect of variations in the environmental conditions can be relevant in this respect and will be investigated in the next section. There, we present the same data but grouped in periods of one month to study their time evolution and the correlation with the ambient variables. As an additional control, we have also verified that the total neutron rates given for each \textit{Phase} (Table~\ref{Tab2}) are fully consistent with the rates obtained independently as weighted-averages of the monthly rates of Section~\ref{evol}. The neutron rates calculated by using the simulated flux curves from Ref.~\cite{Mon_2021} corresponding to different types of rock and concrete (see Section~\ref{an}) reproduce the order of magnitude of our experimental neutron rates (Table~\ref{Tab2}). A detailed characterization of the composition of the concrete in Hall~A is necessary and will be carried out soon.

\begin{table*}[th!]
	\centering
	\caption{Neutron counting rates measured with HENSA in the Hall~A of LSC. The rates measured in each detector during the whole \textit{Phases}~1, 2 and 3 are reported. The last column shows neutron counting rates from a previous measurement \cite{Jordan} done in the Hall~A using the same PE moderators as presently, but the detectors had $P$~=~20 atm and there were other differences in the measurement conditions, as described in the text.}
	\label{Tab2}
		\begin{tabular*}{\textwidth}{@{\extracolsep{\fill}}lllll@{}}
		\hline\noalign{\smallskip}
		 & \multicolumn{1}{l}{\textit{Phase}~1} & \multicolumn{1}{l}{\textit{Phase}~2} & \multicolumn{1}{l}{\textit{Phase}~3} & \multicolumn{1}{l}{Ref. \cite{Jordan}} \\
     Period    & \multicolumn{1}{l}{Oct19 - Feb20} & \multicolumn{1}{l}{Mar20 - Jul20} & \multicolumn{1}{l}{Aug20 - Feb21} & \multicolumn{1}{l}{Jun11} \\
		 Live days & \multicolumn{1}{l}{113} & \multicolumn{1}{l}{135} & \multicolumn{1}{l}{164} & \multicolumn{1}{l}{25.3} \\ \hline
		 Detector & \multicolumn{3}{c}{Neutron counting rate (10$^{-4}$ s$^{-1}$)} & \\ \hline
			E1 & 5.50(15) & 4.07(20) & 4.64(24) &              \\
			E2 & 4.52(14) & 5.38(12) & 4.37(17) & 4.38(20) \\
			E3 & 4.98(11) & 4.75(11) &               & 5.04(21) \\
			E4 & 4.58(11) & 4.28(13) &               & 3.79(19) \\
			E5 & 2.34(8)   & 2.20(9)   &               & 2.33(16) \\
			E6 & 1.38(9)   & 1.39(6)   &               & 1.28(12) \\
			E7 & 0.81(5)   & 0.75(4)   &               & 0.77(10) \\
			E8 & 2.20(8)   & 2.24(7)   &               &               \\
		  E9 & 0.42(4)   & 0.46(3)$^a$ &          &                \\
    \noalign{\smallskip}\hline
		\end{tabular*}
	\\ \raggedright{$^a$ This value includes a correction accounting for the different pressure of E9 in \textit{Phases}~1 and 2 (10 and 20 atm, respectively).}
\end{table*}

In 2011 a first and shorter measurement with $^3$He detectors was carried out at LSC \cite{Jordan}. It took place in the almost empty Hall~A, before the beginning of any scientific experiment at LSC, and lasted 26 days. Six $^3$He counters were employed, with the same PE moderators as for the detectors from E2 to E7 which are listed in the present Table~\ref{Tab1}. It is worth noting that in Ref.~\cite{Jordan} the detectors had a pressure of 20 atm, while in the present measurement $P$ = 10 atm. The six detectors of the previous measurement were positioned in the middle of Hall~A, distributed in a fan-like arrangement on a light structure one meter above ground \cite{Jordan}. To facilitate the comparison with the present results, we show the results from Ref.~\cite{Jordan} in the last column of Table~\ref{Tab2} keeping the present nomenclature for the detectors. The data have not been corrected to account for the different pressure. The measurement conditions were also different. Indeed, the previous data \cite{Jordan} were taken in the middle of an empty hall while currently Hall~A is full of materials from the various experiments and the HENSA setup is closer to the walls. This can influence the neutron rate locally, which is sensitive to changes in the composition of the surrounding environment. Having these differences in mind, there is a general agreement being the values consistent within 1 sigma in most of the cases. 

In the previous measurement a background contribution was observed in the region [$200,800$] keV, extending into the neutron signal region (see Fig.~4 in Ref.~\cite{Jordan}). In HENSA we lowered the $^3$He gas pressure from 20 to 10 atm in order to decrease the sensitivity to $\gamma$ rays, achieving a substantial reduction of this background component, as shown clearly in Figs.~\ref{Fig2} and \ref{Fig3}. The electronic noise has also been decreased by improvements in the electromagnetic compatibility of our electronic setup. Moreover, the addition of the detectors labelled E1, E8 and E9 in Tables~\ref{Tab1} and \ref{Tab2} has improved the sensitivity to thermal and high-energy neutrons \cite{HENSA1,Orr_2021}.

Starting from the neutron rate data of Table~\ref{Tab2} and knowing the HENSA response to neutrons of different energy, one is able to deduce both the integral neutron flux and its spectral distribution. An example of this kind of analysis and obtainable results is shown in Fig.~4 of Ref.~\cite{Orr_2021}, reporting the preliminary results on the neutron flux from the analysis of \textit{Phase}~1 (113 live days). The preliminary value obtained for the total neutron flux in Hall~A of LSC was \mbox{$1.66(2) \times 10^{-5}$ cm$^{-2}$ s$^{-1}$}, in reasonable agreement with the value of the previous measurement \mbox{$1.38(14) \times 10^{-5}$ cm$^{-2}$ s$^{-1}$} \cite{Jordan} considering the different experimental conditions. Improvements on the analysis and detector response as well as the data analysis of \textit{Phases}~2 and 3 are in progress and will be presented in a forthcoming publication \cite{Orrigo}.

\subsection{\label{evol}Long-term evolution of the neutron rate}

\begin{figure*}[t!]
	\centering
		\includegraphics[width=1\columnwidth]{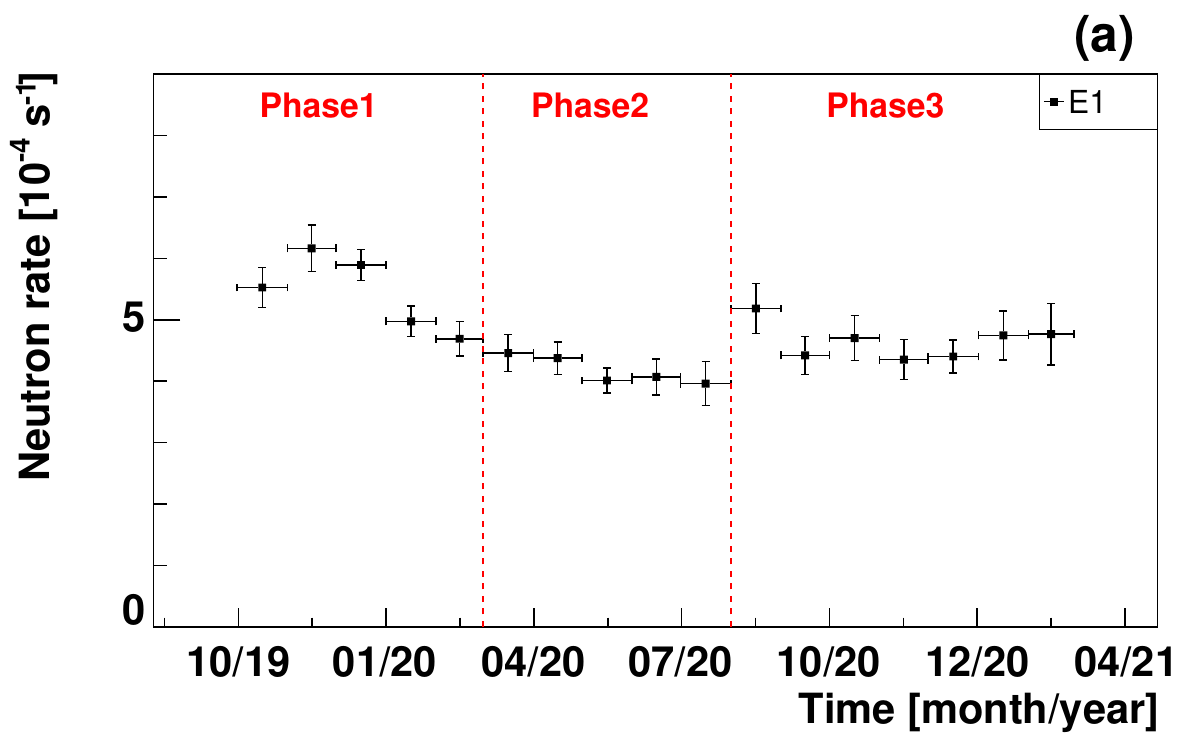}
	  \includegraphics[width=1\columnwidth]{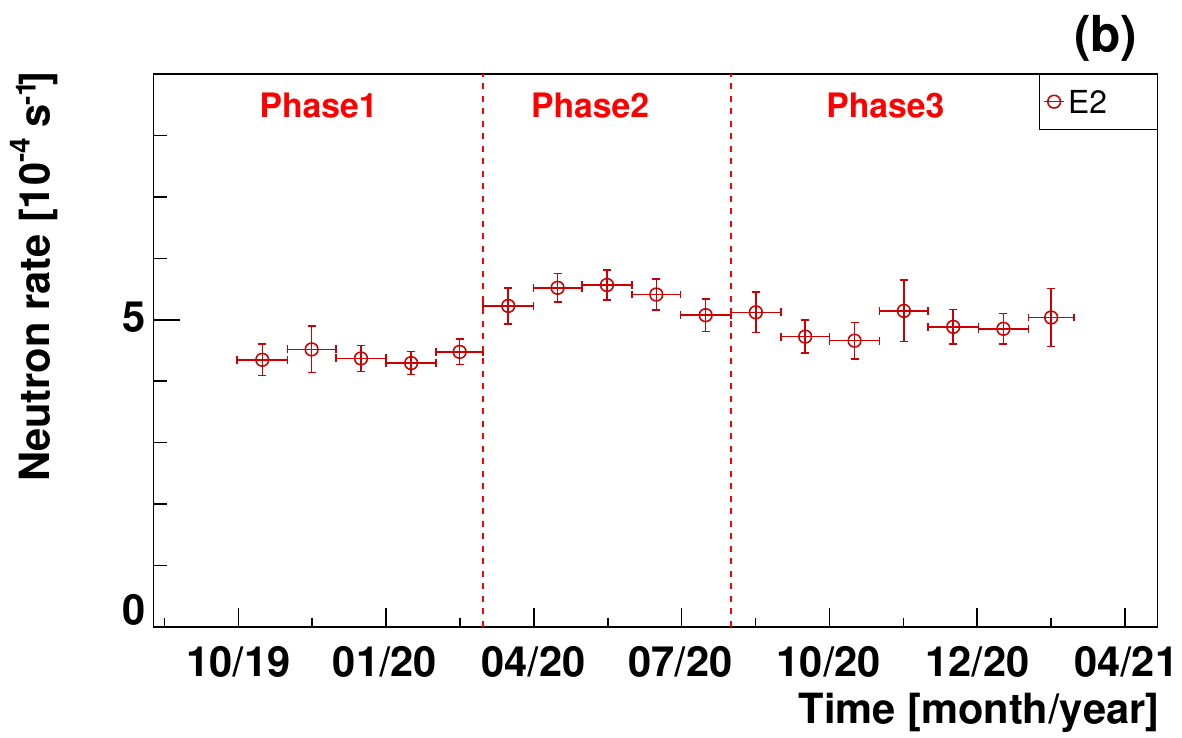}
		\includegraphics[width=1\columnwidth]{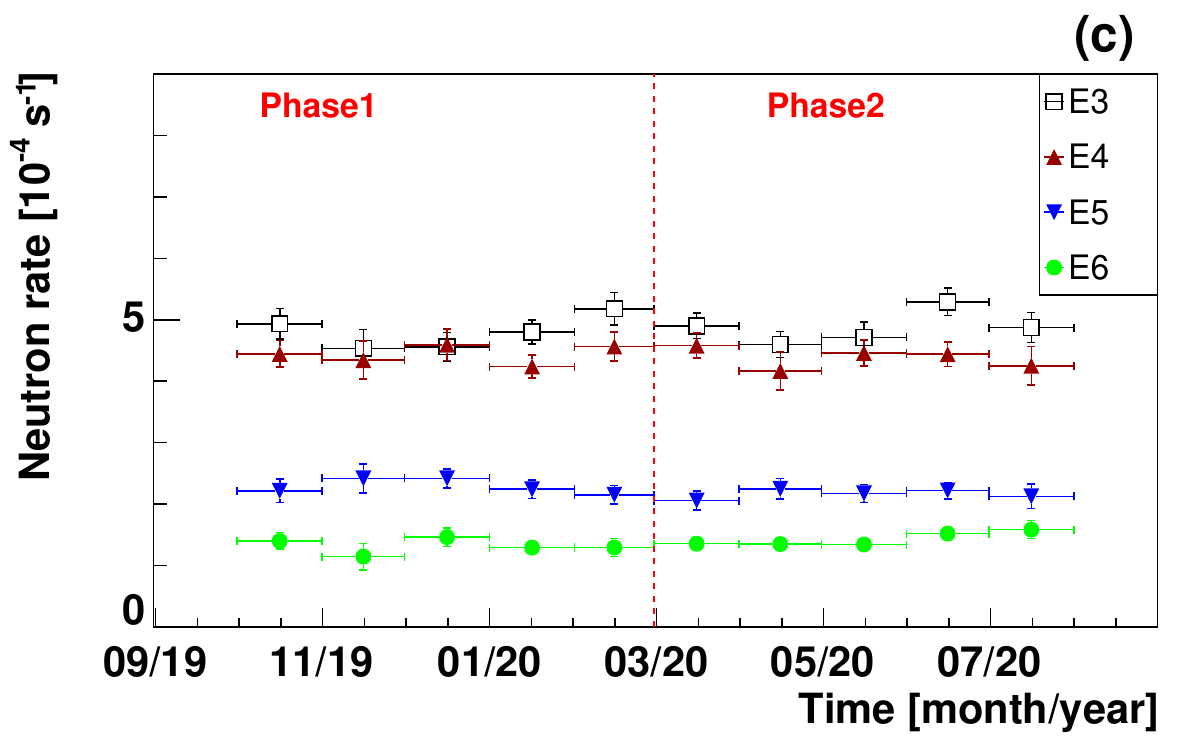}
	  \includegraphics[width=1\columnwidth]{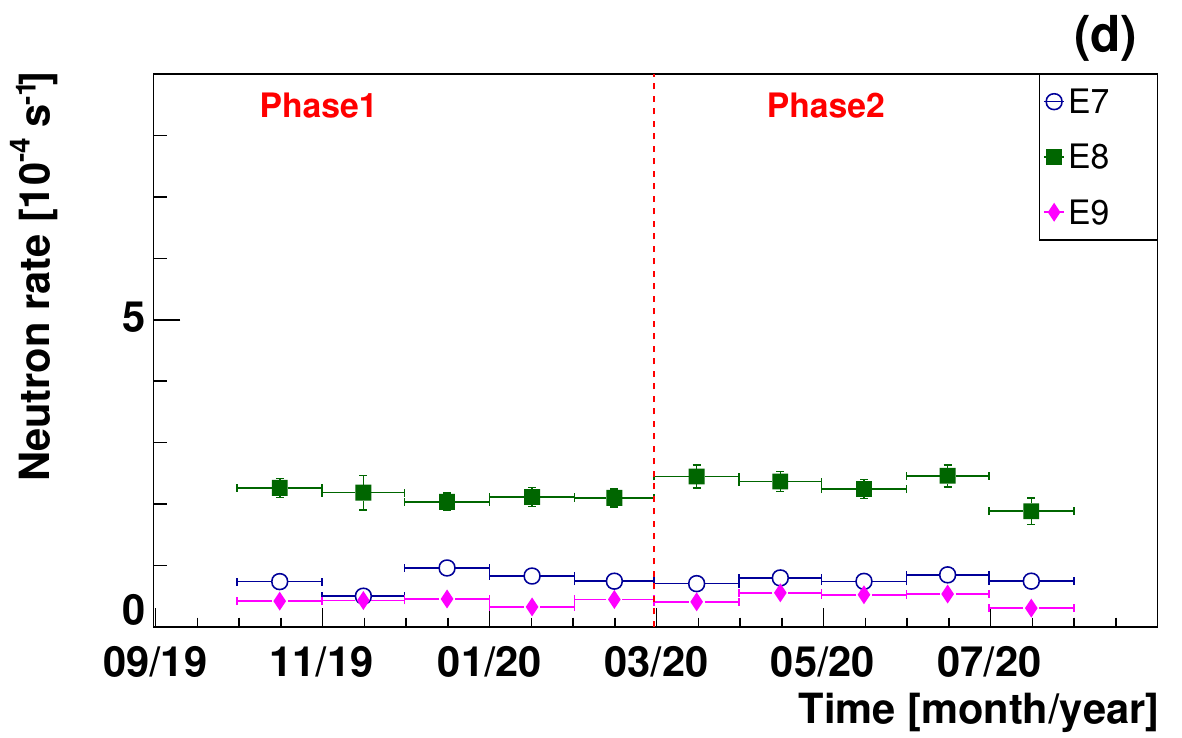}
 	\caption{\label{Fig4} Long-term evolution of the neutron rate observed during our measurement campaign in the Hall~A of LSC in the detectors: (a) E1; (b) E2; (c) E3, E4, E5 and E6; (d) E7, E8 and E9.}  
\end{figure*}

In the present section, the same neutron rate data of Section~\ref{nRate} are presented regrouped in time periods of one month each, in order to investigate their long-term evolution and compare them to the ambient data measured during the same periods. Fig.~\ref{Fig4} shows the long-term evolution of the neutron counting rate measured in each detector (from E1 to E9 according to their sensitivity to increasing neutron energy) during the campaign in Hall~A. In the figure, the time period corresponding to each month is represented by the horizontal error bar. Since the E1 and E2 detectors were employed during all the measurement campaign, the long-term evolution study of the thermal E1 and epithermal E2 rates (Figs.~\ref{Fig4}a and \ref{Fig4}b, respectively) covers 412 live days, corresponding to 17 months. Differently, the detectors from E3 to E9 were not used during \textit{Phase}~3, thus the time evolution of the corresponding rates (Figs.~\ref{Fig4}c and \ref{Fig4}d) refers to the first 248 live days of \textit{Phases}~1 and 2, corresponding to 10 months. 

As shown in Table~\ref{Tab2}, the rates measured in the detectors from E3 to E9 do not change in the different \textit{Phases}. In agreement with that behaviour, the monthly-averaged rates in these detectors (Figs.~\ref{Fig4}c and \ref{Fig4}d) show only variations within the uncertainty and can be regarded essentially as featureless. While the high-energy rate \mbox{($>$1 eV)} is practically flat, the situation is not so straightforward for the thermal and epithermal rates (Figs.~\ref{Fig4}a and \ref{Fig4}b, respectively). Even if they both show some variation, a clear signature of seasonal modulation is not visible across all the measurement period. When restricting to \textit{Phases}~1 and 2 the thermal and epithermal rates look to exhibit some anticorrelation between them (with Pearson linear correlation coefficient of -0.8), however this feature disappears in the \textit{Phase}~3 data (the last seven months in the figures). Another finding is that the thermal rate shows a slightly higher value during the first three months of the campaign. The LSC staff informed us that in December 2019 (third month of our measurement campaign) a clean room made by a metacrilate tent and containing a big PE block was installed in the middle of Hall~A. Since these materials are effective neutron moderators, we think that such a change in the Hall~A environment can be the reason for the observed decrease in the thermal rate. Indeed, the local neutron flux is very sensitive to variations in the composition of the surrounding environment \cite{Best_2016}. It is also important to investigate the influence of the environmental conditions, which we do in the next section.

Our data show that the energy distribution of the neutron background can be affected by modifications in the hall configuration. These effects are difficult to be foreseen, thus it is important to keep on continuous monitoring of the neutron background with spectral sensitivity in order to assess potential impacts on the experiments in operation inside the facility.

\subsection{\label{envr}Long-term evolution of the environmental data}
The LSC is equipped with AlphaGuard monitors of the ambient variables ($^{222}$Rn concentration, air temperature $T$, air pressure $P$ and relative humidity $H$), located in various positions inside the experimental halls.

\begin{figure*}[t!]
	\centering
	  \includegraphics[width=1\columnwidth]{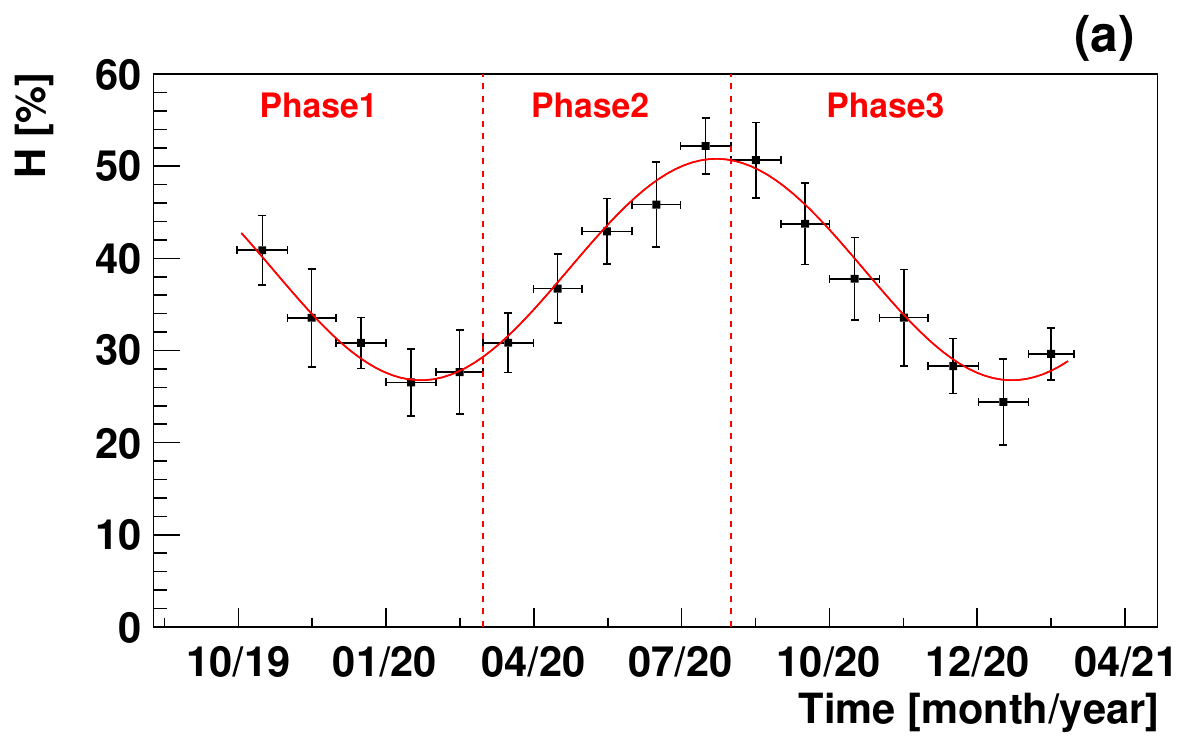}
	  \includegraphics[width=1\columnwidth]{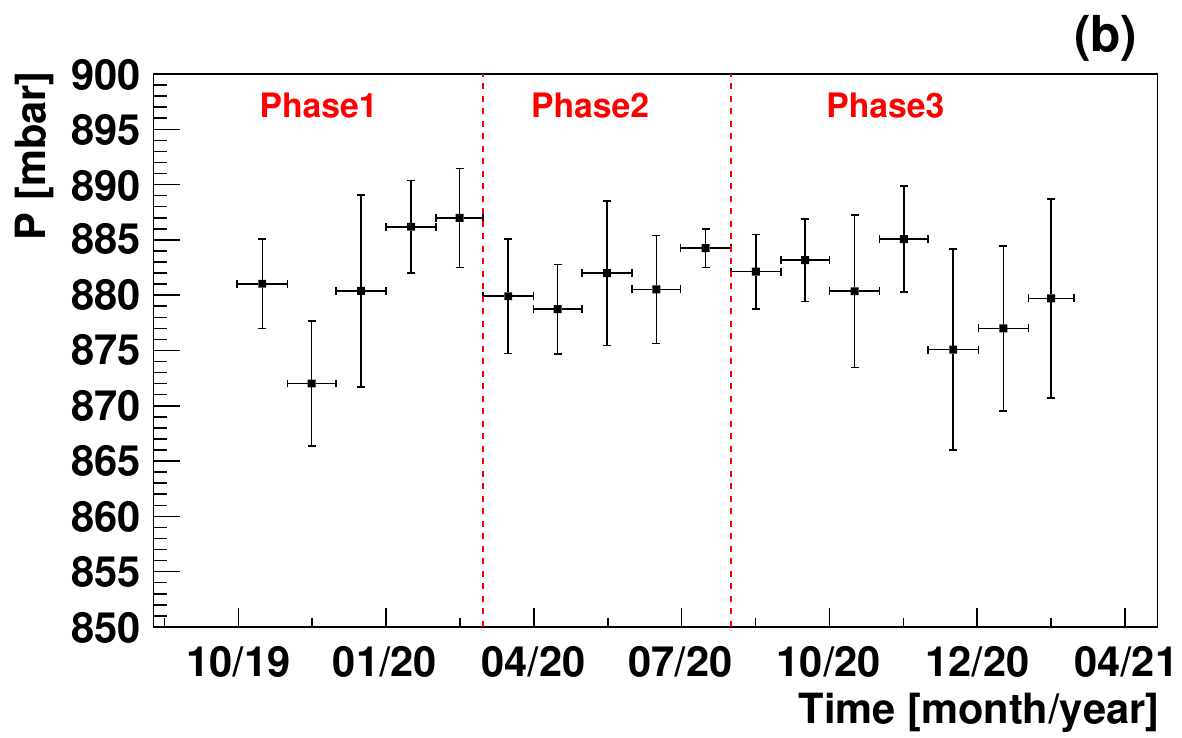}
	  \includegraphics[width=1\columnwidth]{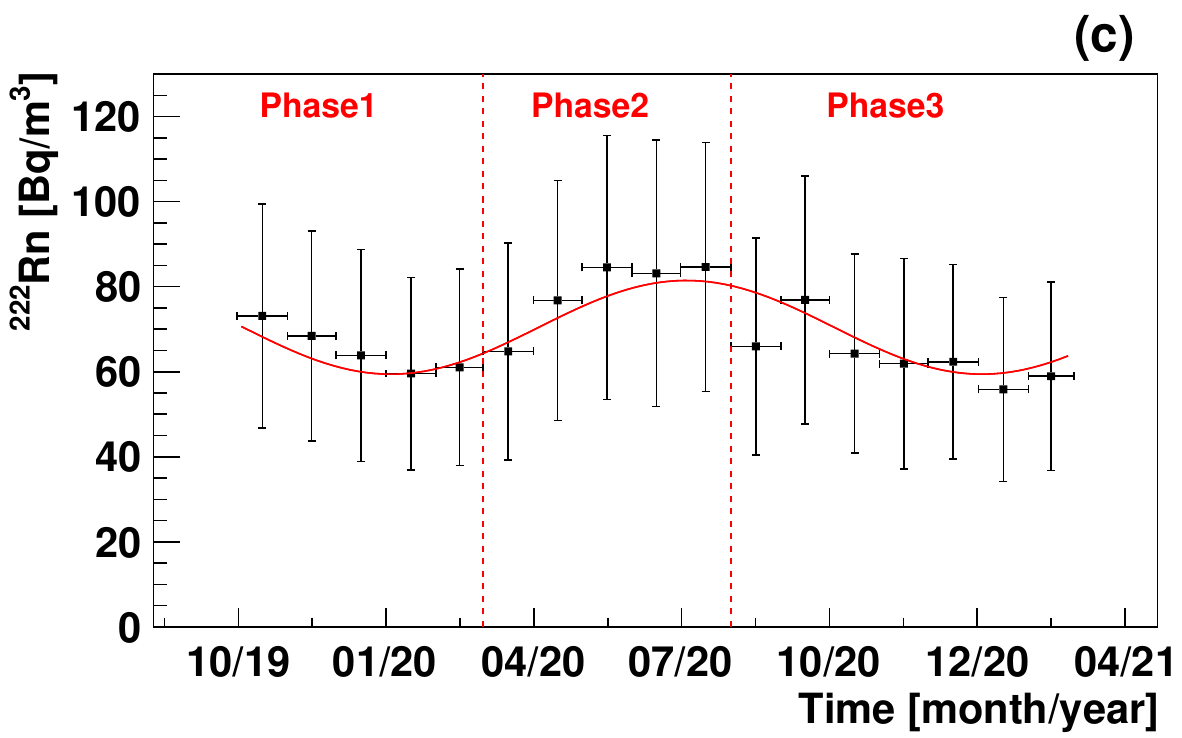}
	  \includegraphics[width=1\columnwidth]{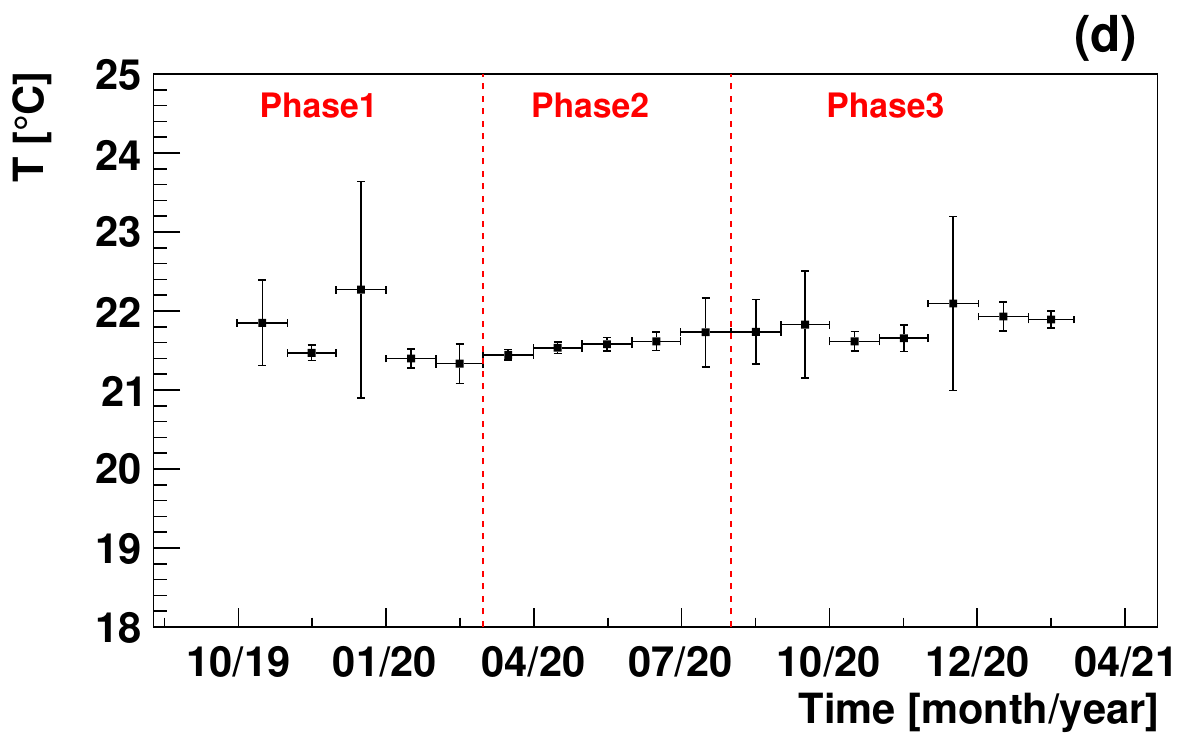}
 	\caption{\label{Fig5} Long-term evolution of the environmental data in the Hall~A of LSC, observed in the same period of our measurement campaign. (a) Relative humidity. (b) Air pressure. (c) Radon concentration. (d) Air temperature.}
\end{figure*}

We have analysed the data of the AlphaGuard located closer to HENSA in Hall~A during the same period of our measurement campaign, covering a total of 17 months (from October 2019 to end of February 2021). The instrument provides 10-minute averaged ambient data. In order to facilitate the comparison with the measured neutron rates (Fig.~\ref{Fig4}), the environmental data are represented in Fig.~\ref{Fig5} as follows. For every ambient variable, the mean value of the values observed during each month is shown. The associated error bar represents the standard deviation of the distribution of values in that given month. The horizontal error bar visualizes the correspondent time interval.  

A seasonal variation is observed in the relative humidity (Fig.~\ref{Fig5}a), which is described by the function:
\begin{equation}
  y(t)=A \cos [ 2\pi (t-\phi)/T ] + C \,,
  \label{Eq3}
\end{equation}
where $A$ is the amplitude of the modulation over the constant component $C$, $T$ is the period and $\phi$ is the phase. The fit shown in Fig.~\ref{Fig5}a, with a period fixed to one year, gives a maximum amplitude in summer, on July 23, with a modulation amplitude of 12(1)\% over a constant baseline of 39(1)\%.

A similar annual variation is found in the $^{222}$Rn concentration (Fig.~\ref{Fig5}c). The fit shown in the figure gives a maximum amplitude on July 4, with a modulation amplitude of 11(9) Bq/m$^3$ over a constant baseline of 70(7) Bq/m$^3$. It should be noticed that the $^{222}$Rn concentration is affected by larger uncertainties arising from the larger variability of the values observed in each month. Such a sizeable variance, in turn, is due to the fact that conventional radon-meters measure the radon concentration in air and so they are quite sensitive to the air movements. Nevertheless, the modulation effect is still visible. Moreover, the $^{222}$Rn concentration and relative air humidity show a positive correlation, with Pearson linear correlation coefficient of 0.8.

Differently from the correlated annual modulation observed in $H$ and $^{222}$Rn, the air pressure and temperature (Figs.~\ref{Fig5}b and \ref{Fig5}d, respectively) have a rather featureless behaviour with mean values of 883(3) mbar and \mbox{21.6(2) $^\circ$C}, respectively. The temperature at LSC is controlled by an air conditioning system which keeps it constant (only in very few occasions it was not working properly). The ventilation system at LSC is partially connected to the outside, indeed the 50\% of air inside the laboratory is recirculated while the other 50\% is introduced from outside. This feature can explain the seasonal variation that we observe in humidity and radon, since during winter there is a drier weather.

It is worth to notice that the behaviour of the environmental variables needs to be monitored, in connection with the neutron rate measurements, because it can be quite different in diverse UG laboratories. We discuss some examples below showing that this behaviour depends on the particular conditions existing in the given UG laboratory. For example, the Large Volume Detector (LVD) experiment located in the Hall~A of LNGS (3600 m.w.e.) observed there that the temperature and humidity were rather featureless, while the radon concentration had a seasonal variation of 4\% with a maximum at the beginning of September, when the ambient rock is most strongly permeated with underground water \cite{Aga_2019}. No neutron detectors were used in Ref.~\cite{Aga_2019}, hence there is no information available on the neutron rate.

We found the pressure quite stable at LSC and not correlated to the seasonal radon variation. Again, this feature can be different in other UG laboratories or when the conditions change on smaller time scales. For example, the forced ventilation system of LNGS produced daily variations in the radon concentration due to the excess of air pressure generated when closing the hall gates at night, reducing the radon emanation from the ambient rock \cite{Aga_2019}. In absence of forced ventilation, sporadic increases in the thermal neutron rate were observed UG at Moscow State University (25 m.w.e.) \cite{Stenkin_2017} when the atmospheric pressure was decreasing: the so-called \textit{radon~barometric~pumping~effect}, i.e., an increase in the radon gas advection as a consequence of the decreasing pressure.

At the Baksan Neutrino Observatory ($\approx$1000~m.w.e.) Ref.~\cite{Alekseenko_2010} reported a seasonal variation in the relative humidity and air temperature with maximum in summer and a similar variation of $\approx$2\% in the amplitude of the thermal neutron rate (but stating that the detector own background was not subtracted). At DULB-4900 (4900 m.w.e.) the same authors observed a seasonal variation in the absolute air humidity with maximum in summer and a correlated variation of $\approx$5\% in the thermal neutron rate \cite{Alekseenko_2017}. The increase of the thermal rate with the humidity was attributed to a better moderation of the neutrons both in the humid air and when exiting from a humid rock as well as higher albedo of the neutrons incoming into the rock. These studies employed $^6$LiF+ZnS(Ag) scintillators, capable to extract information for the thermal rate only.

Our measurement is able to provide data on both the thermal and epithermal rates as well as the high-energy ones ($>$1 eV up to 0.1 GeV and beyond). In Section~\ref{envr} we have shown that above 1 eV the rates are essentially flat as a function of time. Even if we observe some kind of variation in both the thermal and epithermal rates, there is no clear correlation with the ambient variables across all the duration of the measurement campaign. When restricting to the \textit{Phases}~1 and 2 periods only (excluding the first three months according to Section~\ref{evol}) the thermal rate exhibits anticorrelation (Pearson coefficient of -0.94) with both the humidity and radon concentration. However this finding could be incidental because this feature is absent in \textit{Phase}~3. At the same time, it is not clear if there might have been some not monitored change in the Hall~A conditions in \textit{Phase}~3. Therefore, further monitoring and investigation of the interplay between neutron rate and environmental variables is necessary. 

\section{\label{concl}Conclusions}
We have carried out the first long-term measurement (412 live days) of the neutron rate in an UG laboratory with a $^3$He-array spectrometer sensible to a broad range of neutron energies. Few studies of neutron background in UG laboratories, with sensitivity to the thermal neutron rate only, existed in the literature before the present measurement, and none of them at the LSC. We have studied the time evolution of the neutron rate at the LSC not only in the thermal range, but also in the epithermal and high-energy ranges. We have found that the high-energy rate is featureless. Since high-energy neutrons constitute a potentially dangerous background for astroparticle experiments, our observation of a constant high-energy rate is a first significative information. On the other hand, the thermal and epithermal rates do show a variation. No clear signature of seasonal variation is found in the thermal rate across all the measurement campaign. The correlation with the ambient variables (radon concentration, air temperature, pressure and humidity) has been investigated, but the conclusion is not straightforward. We have also observed that the rates can be affected by the conditions (external activity, addition/removal of materials, etc.) inside the experimental hall. The complex interplay of meteorological variables, conditions in the hall and neutron rate points out the needs for long-term measurements devoted to monitor and characterize the rate at the experimental location. These effects are not easily predictable, therefore it is of paramount importance to carry on a continuous monitoring of the neutron background with spectral sensitivity to estimate the possible influence on the low-rate experiments running at LSC.

\begin{acknowledgements}
This work was supported by the Spanish Grants No.~PID2019-104714GB-C21, PID2019-104714GB-C22, RTI2018-098868-B-I00, FPA2017-83946-C2-1-P and FPA2017-83946-C2-2-P (MCIN, MCIU, MINECO /AEI/FEDER). We acknowledge the support of the Generalitat Valenciana Grant No.~PROMETEO/2019/007. We are grateful to Laboratorio Subterr{\'a}neo de Canfranc for hosting the HENSA spectrometer, for the support received by the LSC personnel during the measurement campaign in Hall~A and for providing us with the environmental data.
\end{acknowledgements}

\bibliography{references}

\end{document}